\documentstyle[prl,aps,epsf,floats]{revtex}

\begin{document}
\draft

\twocolumn[\hsize\textwidth\columnwidth\hsize\csname @twocolumnfalse\endcsname
\title{Orientational dependence of current through molecular films} 
\author{P.E. Kornilovitch and A.M. Bratkovsky}
\address{
Hewlett-Packard Laboratories, 1501 Page Mill Road, Palo Alto, 
California 94304
}

\date{ February 16, 2001 }
\maketitle

\begin{abstract}

We study the current through molecular films as a function of
orientation  of the molecules in the film with respect to electrodes.
It may change by more than an order of magnitude,
depending on the  angle between the axis of the molecules and the
normal to the electrode. This is a 
consequence of a strong directional character of $p$-orbitals that determines 
the conductance through the molecules.  We demonstrate this result 
on an exactly solvable model, and present the 
calculations for two  different experimentally accessible molecular films
sandwiched between gold  electrodes.
 
\end{abstract}

\vskip2pc] \narrowtext


The studies of organic molecules as possible electronic components and
devices have grown out of the initial suggestion by Aviram and Ratner \cite
{Aviram} into a very active area of research, see e.g. \cite
{Aviram98,RatnerRev99}. One of the most important issues in this problem is
the role of the molecule-surface interface. The geometry of the contacts 
\cite{Yaliraki98,Kirczenow,Pantelides}, the type of binding \cite
{Kirczenow,Roitberg,Onipko}, and the molecule-electrode distance \cite
{Kirczenow,Magoga,Yaliraki} were all found to affect the conductance
significantly. In this paper we perform a model study of the effect of
mutual orientation of the molecules and the electrodes on conductance. We
find that the orientational dependence of the current through molecular
films is much stronger than reported previously in Ref.\cite{Yaliraki}, and
it should be taken into consideration in both interpreting experimental data
and designing possible moletronic devices.

The simple argument in favor of a substantial orientational dependence of
the conductance is the large anisotropy of the molecule-electrode coupling.
In most molecules that have been studied experimentally or theoretically so
far, the conduction is due to $\pi $-conjugated molecular orbitals (MOs).
Such MOs extend over the whole molecule and facilitate the transport of
electrons between the two electrodes. Since those MOs are made mostly of
anisotropic $p$ atomic orbitals, the overlap of the conducting MOs with
electrode wave functions strongly depends on the angle between the main axis
of the molecule and the surface normal. In general, we expect the overlap
and the full conductance to be maximal when the lobes of the $p$-orbital of
the end atom at the molecule are oriented perpendicular to the surface, and
smaller otherwise, as dictated by the symmetry. An estimate of the effect
can be made from the general properties of the $p$-orbitals. The overlap
integrals of a $p$-orbital with orbitals of other types differ by a factor
about 3 to 4 for the two orientations \cite{Harrison}. Since the conductance
is proportional to the square of the matrix element, which contains a
product of two metal-molecule hopping integrals, the total conductance
variation with overall geometry may therefore reach two orders of magnitude,
and in special cases be even larger. This conclusion would be only valid
when the electronic structure of the molecule and the molecule-electrode
distance does not change significantly while its orientation is being
changed. One can anticipate this to apply when the coupling to the electrode
is weak, as might be the case in some experimental setups. The calculations 
\cite{Sellers} suggest that there are attachements of particular molecules
with very little resistance to the rotation with respect to metallic
surface. The present effect (change of the geometry of the metal-molecule
contact) should be distinguished from other geometrical effects caused by 
{\em intra}molecular conformation changes. One possibility, discussed in
Ref. \cite{magoga99}, is to rotate a segment of a $\pi $-conjugated
molecular wire. That would break the conjugation and result in decreased
conductance. Another would be to rotate an active part of the molecule with
respect to the rest of it to effectively change the current path,  which
leads to significant changes in resistance of the effective circuit\cite
{moresco01}.

In order to illustrate the geometric effect on current we shall first
consider a simple model shown in Fig.~\ref{fig2} (top panel). A molecular
film is sandwiched between two three-dimensional electrodes, which are
assumed to have a simple cubic structure with one $s$-orbital per site and
the onsite energy $\epsilon _{s},$ which we take as an energy origin. We
assume, without loss of generality, that the band in the leads is
half-filled, so that the Fermi level $E_{F}=\epsilon _{s}=0.$ The hopping
integral between the neighbors is $-t_{s}.$ The ``molecules'' have only two $%
p$ $\pi-$type atomic orbitals with the onsite energies $\epsilon _{p}$ and
hopping between those $\pi -$orbitals is $-t_{\pi } $. Those orbitals model $%
\pi $-conjugated bands of organic conductors. The molecules are allowed to
rotate about the bottom ``atom'' and thereby change the angle $\theta $
between the main axis of the molecules and the surface normal. As $\theta $
changes, the $p$-orbitals remain oriented perpendicular to the molecular
``backbone''. Because of that, the matrix element between a molecular $p$%
-orbital and an electrode $s$-orbital varies as $t_{sp}\sin {\theta }$,
where $t_{sp}$ is a constant ($sp-$ hopping matrix element between the
molecule and the electrode). The molecule-electrode bond lengths do not
change with $\theta $, and the surfaces of the two electrodes always remain
parallel. Each molecule is assumed to have one electron per atom. The
distance between the lowest unoccupied MO (LUMO) and highest occupied MO
(HOMO), or the ``HOMO-LUMO'' gap, is $E_{{\rm LUMO}} - E_{{\rm HOMO}%
}=2t_{\pi }.$ The middle of the gap is aligned with the chemical potential
of the leads at zero bias voltage $V=0$ (meaning that $\epsilon
_{p}=\epsilon _{s}=0$).

\begin{figure}[t]
\epsfxsize=7.5cm
\epsffile{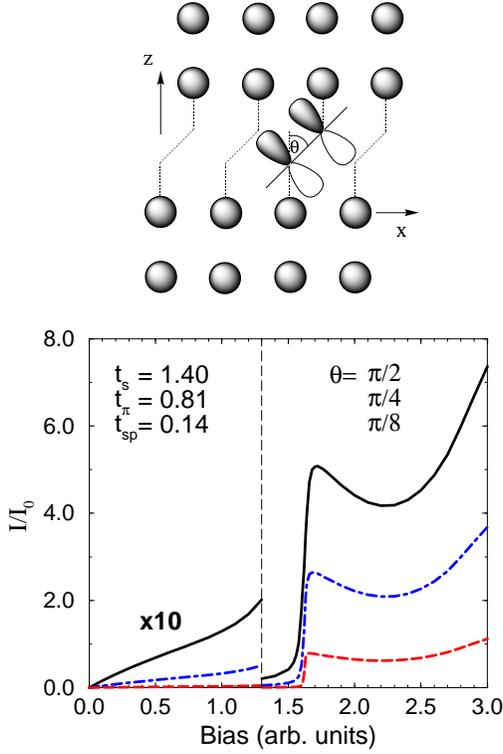}
\caption{ Top panel: the molecular film between two leads with simple cubic
structure and one $s$-orbital per site. Only one molecule of the film is
shown explicitly. Due to the symmetry of coupling to the electrodes, the
current is zero for the normal orientation ($\protect\theta=0$) and maximal
for the parallel orientation ($\protect\theta=\protect\pi/2$) of the
molecules. Bottom panel: current per surface unit cell versus bias voltage
for the tilt angles $\protect\theta=\protect\pi/2$, $\protect\pi/4$, and $%
\protect\pi/8$, with the hopping parameters $t_{s}$, $t_\protect\protect\pi$%
, and $t_{sp}$ in arbitrary units ($t_0$). Bias is in units of $t_0/|e|$ ,
and the parameter $I_0 = 10^{-3}t_0\frac{2e}{h}$. }
\label{fig2}
\end{figure}

The current through the film at low temperatures is given by a standard
expression \cite{Landauer,newGF} 
\begin{equation}
I=\frac{2\left| e\right| }{h}\int_{E_{F}-\left| e\right| V/2}^{E_{F}+\left|
e\right| V/2}dE\sum_{\vec{k}_{\parallel }}^{{\rm open}}\left| t(E,\vec{k}%
_{\parallel })\right| ^{2},
\end{equation}
where the summation goes over the surface Brillouin zone of the lead for
open channels, and the transmission coefficient is found from solution of
the scattering problem\cite{newGF}, 
\begin{equation}
t(E,\vec{k}_{\parallel })=iD^{-1}\sqrt{v_{zL}v_{zR}}t_{\pi
}t_{sp}^{2}e^{i(k_{zL}+k_{zR})}\sin ^{2}\theta ,  \label{eq:trexact}
\end{equation}
where 
\begin{eqnarray}
D &=&(E^{2}-t_{\pi
}^{2})t_{s}^{2}+Et_{s}t_{sp}^{2}(e^{ik_{zL}}+e^{ik_{zR}})\sin ^{2}\theta 
\nonumber \\
&+&e^{i(k_{zL}+k_{zR})}t_{sp}^{4}\sin ^{4}\theta ,  \label{eq:D}
\end{eqnarray}
and we assume $\hbar =1.$ Then, the band velocities are $v_{zL(R)}=2t_{s}%
\sin k_{zL(R)},$ with $k_{zL(R)}$ the z-components of the momenta in the
leads, which are found from 
\begin{equation}
2t_{s}\cos k_{zL(R)}=-E+(-)\frac{\left| e\right| V}{2}-2t_{s}(\cos
k_{x}+\cos k_{y}).  \label{eq:kz}
\end{equation}
It is instructive to analyze the transmission in the limiting case of weak
coupling to the electrodes, $t_{sp}/t_{\pi }\ll 1.$ The exact transmission
probability, Eq.~(\ref{eq:trexact}), acquires the resonant form for zero
bias voltage, $V=0,$ when $k_{zL}=k_{zR}$: 
\begin{equation}
\left| t(E,\vec{k}_{\parallel })\right| ^{2}\approx \sum_{r=1,2}\frac{\Gamma
^{2}/4}{\left( E-E_{r}\right) ^{2}+\Gamma ^{2}/4},  \label{eq:BW}
\end{equation}
with 
\begin{eqnarray}
E_{1(2)} &=&+(-)t_{\pi }-\Delta ,  \label{eq:Er} \\
\Delta &=&\Delta \left( E,\vec{k}_{\parallel };V\right) =\left(
t_{sp}^{2}/t_{s}\right) \cos k_{z}\sin ^{2}\theta , \\
\Gamma &=&\Gamma \left( E,\vec{k}_{\parallel };V\right) =\left(
t_{sp}^{2}/t_{s}\right) \sin k_{z}\sin ^{2}\theta ,
\end{eqnarray}
where $\Delta ,\Gamma \ll E_{1}-E_{2}=2t_{\pi }$ is the shift of the
resonant levels, and their width, respectively. For non-zero bias the
transmission probability is characterized by different line widths for
hopping to the left (right) lead $\Gamma _{L}\left( \Gamma _{R}\right) ,$
and Eq.~(\ref{eq:BW})\ will have a general form $\sum_{r=1,2}\frac{\Gamma
_{L}\Gamma _{R}}{\left( E-E_{r}\right) ^{2}+\Gamma ^{2}/4}$ with $\Gamma
=\Gamma _{L}+\Gamma _{R}.$ If we were to neglect the dependence of $\Gamma $
on energy and $\vec{k}_{\parallel },$ we would obtain 
\begin{equation}
I\approx \left\{ 
\begin{array}{cc}
\frac{e^{2}}{h}\frac{\Gamma ^{2}}{t_{\pi }^{2}}V\propto \sin ^{4}\theta , & 
\left| e\right| V\ll E_{{\rm LUMO}}-E_{{\rm HOMO}} \\ 
\frac{8\pi \left| e\right| }{h}\frac{\Gamma _{L}\Gamma _{R}}{\Gamma
_{L}+\Gamma _{R}}\propto \sin ^{2}\theta , & \left| e\right| V>E_{{\rm LUMO}%
}-E_{{\rm HOMO}}.
\end{array}
\right.  \label{eq:I2case}
\end{equation}
The low bias case corresponds to sub-resonant tunneling inside the HOMO-LUMO
gap, whereas the case of large bias involves the tunneling through HOMO and
LUMO. There is a crossover in the angular dependence of the current through
the molecule from $\sin ^{4}\theta $ to $\sin ^{2}\theta $ with the bias.
The current saturates at large bias (with further changes due to the density
of states in the leads) and its angular dependence becomes less steep. This
behavior is confirmed by the results of exact calculations shown in Fig. \ref
{fig2}. It shows a dependence of current on the angle, which is stronger at
small biases. The maximum of the current corresponds to $\theta =\pi /2,$
i.e. to the {\em horizontal} orientation of the molecule in the film
(backbone of the molecule parallel to the surface of the electrodes). Exact
results also show the non-monotonous behavior of the current with the bias,
when it is comparable to HOMO-LUMO gap, which comes from the complex bias
dependence of $k_{zL(R)}$ through the dispersion relations (\ref{eq:kz}).

The formula (\ref{eq:BW})\ is, obviously, a particular form of the
Breit-Wigner formula and it should apply whenever the width of the
resonances is much smaller than the separation between them. Thus, the
results, Eq.~(\ref{eq:I2case}), are more general and should qualitatively
apply in more complex molecules too, if the overlap between resonances is
not substantial, although it is likely to be violated at high bias voltages.

\begin{figure}[t]
\epsfxsize=7.5cm
\epsffile{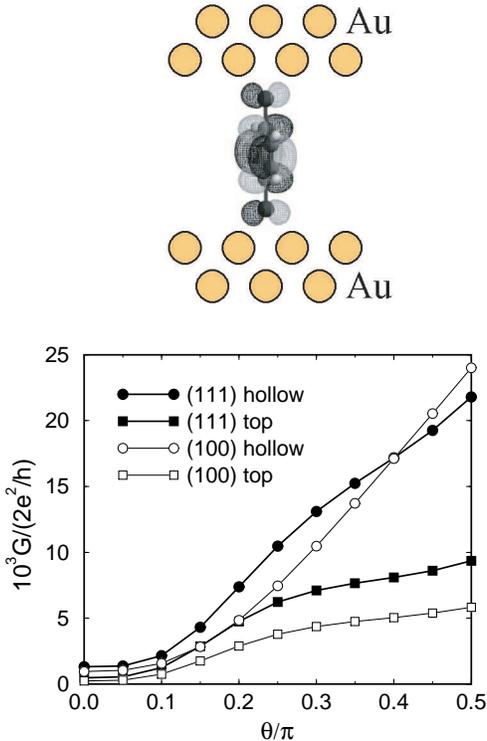}
\caption{ Top panel: schematic of benzene-(1,4)-dithiolate molecules
positioned between two gold electrodes. Also shown is the LUMO of the free
molecule found from the density functional theory. Bottom panel: conductance
as a function of the tilt angle $\protect\theta $. }
\label{fig4}
\end{figure}

We illustrate the angular dependence on two types of molecules,
benzene-(1,4)-dithiolate (-S-C$_{6}$H$_{4}$-S-) and $\alpha ,\alpha ^{\prime
}$-xylyl-dithiolate (-S-CH$_{2}$-C$_{6}$H$_{4}$-CH$_{2}$-S-) placed between
two gold electrodes. Both of those systems were studied experimentally \cite
{Reed,Dorogi} and theoretically \cite
{Kirczenow,Pantelides,Roitberg,Onipko,Yaliraki,Hall,Datta}, but mostly
assuming a particular symmetric position of the molecules with respect to
the surface. Those molecules attach strongly to the gold substrate by thiol,
-S-, end groups, which form a chemical bond with Au \cite{Reed,Dorogi}. The
strong bonding might well mean a substantial change in the electronic
structure of the molecule, because of a possible charge flow to/from the
molecule bonded to a metal (electron reservoir)\ due to the difference in
their ionization potentials. Therefore, the following calculation is rather
the model study to illustrate the orientation effect, which may not be as
strong in a particular setup\cite{Reed,Dorogi} compared to our calcualted
effect. It is clear, however, that the effect exists and may be large in
situations when the coupling to electrodes does not strongly perturb the
molecule itself.

In order to compute the conductance, we follow the same general procedure of
Ref.~\cite{newGF}. The gold electrodes have been described by a tight
binding model with nine $s-$, $p-$, and $d-$orbitals per each Au atom with
the parameters from Ref.~\cite{Papa}. The equilibrium molecular geometry is
found by the total-energy density functional minimization \cite{Spartan}.
The tight-binding parameters for the molecules and molecule-lead interfaces
have been taken from the solid-state table of elements \cite{Harrison}.

Consider the benzene-(1,4)-dithiolate film. We first place the molecules
perpendicular to the gold electrodes, as shown at the top of Fig.~\ref{fig4}%
. Also shown is one of the conducting orbitals, the LUMO. For this
orientation the conductance should be small because the $p$-orbitals, that
constitute the LUMO (and the HOMO, which also conducts), are parallel to the
surface and the overlap is the smallest. Next, we allow the molecules to
rotate about one of the end sulfur atoms while keeping the two surfaces
parallel and both sulfur-gold bond lengths constant. The molecules are
rotated in such a way that the lines through pairs of hydrogen atoms, that
are symmetric with respect to molecules' main axis, remain parallel to the
gold surfaces. As the angle $\theta $ between the molecules and the surface
normal increases, so does the overlap of the sulfur $p$-orbitals with the
gold orbitals. Thus, we expect the conductance to grow with $\theta $.

Our numerical results are shown in Fig.~\ref{fig4}. We have studied two
different surfaces, Au(111) and Au(100), and two binding schemes, the
``on-top'' and the ``hollow'' positions. In the first case, the sulfur atoms
are positioned directly on top of the gold atoms at a distance of 2.39 \AA 
\cite{Sellers}. In the hollow case, the sulfurs are in symmetric positions
with respect to three gold atoms of the Au(111) surface, and with respect to
four gold atoms of the Au(100) surface. In all four cases, we found the
conductance to be a steadily increasing function of $\theta $. The ratio $G(%
\frac{\pi }{2})/G(0)$ is the highest (=26) in the Au(100)-hollow case and
lowest (=17) in the Au(111)-hollow case. The shapes of all curves are
qualitatively similar, while the hollow position is 2 to 3 times more
conductive than the top position. Thus we observe a large orientational
effect, which magnitude is rather insensitive to the electronic structure
and geometry of the electrodes. This finding corroborates our main argument
that the orientation of molecular $p$-orbitals with respect to the
electrodes is an important factor in the problem. 
\begin{figure}[t]
\epsfxsize=7.0cm
\epsffile{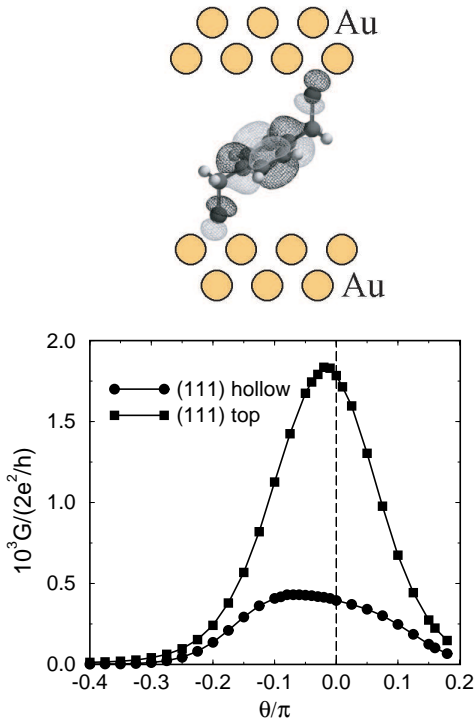}
\caption{ Top panel: schematic of $\protect\alpha ,\protect\alpha ^{\prime }$%
-xylyl-dithiolate molecules placed between two gold electrodes. Also shown
is the LUMO of the free molecules computed within the density functional
theory. Bottom panel: conductance of the film absorbed on the Au(111)
surfaces. }
\label{fig6}
\end{figure}

Next, we consider another molecule, $\alpha, \alpha^{\prime}$%
-xylyl-dithiolate, see Fig.~\ref{fig6}. The $\theta=0$ position is chosen to
be the one with the S-C bond perpendicular to the gold surface. Then, again,
the angle between the S-C bond and the surface normal is systematically
changed in both directions, while keeping the gold-sulfur bond lengths
constant and the two electrode surfaces parallel. Since this molecule is
asymmetric with respect to the S-C bond, the allowed interval of the tilt
angle is also asymmetric, $-0.4 \pi < \theta < 0.2 \pi$ (positive angle
corresponds to the clockwise rotation of the molecules). We have studied the
Au(111) surface and the two binding positions of sulfur, the on-top and the
hollow. The computed conductance is shown at the bottom of Fig.~\ref{fig6}.
It is maximal at small $\theta$ and falls off upon tilting in either
direction. Such a behavior is readily understood if the spatial structure of
the conducting MOs is taken into account. In Fig.~\ref{fig6} one of those
orbitals, the LUMO, is shown. The hybridization of the sulfur $p$-orbitals
with the rest of the molecule has led to the wave function having a
preferential direction from the sulfur atoms towards the ring. Still, the
wave function retains its $p$-character so that a strong anisotropy of the
LUMO-gold overlap is expected. Note that, according to Fig.~\ref{fig6}, the
precise direction of the wave function is in slight misalignment with the
S-C bond, which is chosen as $\theta=0$. The maximum overlap, and
consequently maximum conductance, is therefore expected for a small negative 
$\theta$, precisely the behavior we obtained from the full numerical
treatment of the scattering problem. This detail indicates the adequacy of
the present tight-binding parameterization.

The present results show that the orientational dependence
``molecule-electrode'' of the conductance is a large effect if the
conducting orbitals are extended $\pi $-orbitals, and the contact with
electrode does not strongly perturb the molecule itself. The effect derives
from the strong directional character of $\pi $-orbitals. We have also
studied several other systems of different complexity, both analytically and
numerically, and obtained the results consistent with this simple picture.
It is usually assumed that the geometry in molecular transport experiments
is fixed, but there may be circumstances when the geometry can be changed.
Firstly, for some molecule-electrode pairs there may be more than one
possible chemisorption mode. For instance, for thiolates on gold it was
found in Ref.~\cite{Sellers} that the hollow and on-top binding
configurations are close in energy but have very different surface-S-C bond
angles, $\sim 180^{\circ }$ and $\sim 105^{\circ }$, respectively.
Consequently, the molecules may change the configuration with temperature
and/or under other external factors. According to our calculations, this may
lead to a significant change of conductance. Secondly, the tilt angle in
self-assembled monolayers or Langmuir-Blodgett films, used in moletronics
studies, is never exactly zero \cite{Tour} and may depend on the
temperature, pressure, and/or other parameters of preparation. Thirdly, the
angle may change during the measurement process. For instance, in STM probes
of single molecules \cite{Weiss} the position of the probing tip is not
fixed with respect to the molecule and may record a distribution of
conductance if the latter is strongly angle-dependent. Finally, a finite
temperature always results in geometry fluctuations that in turn lead to
conductance fluctuations. Thus the present mechanism may be one of the
sources of the temperature dependence of the conductance.

We thank E.~Emberly, A.~Onipko, M. Ratner, R.S. Williams, and our colleagues
at Quantum Science Research group, Hewlett-Packard Labs, for useful
discussions. The work has been partly supported by DARPA.



\end{document}